%% file: main.tex
\documentclass[conference]{IEEEtran}
\IEEEoverridecommandlockouts

\usepackage{cite}
\usepackage{amsmath,amssymb,amsfonts}
\usepackage{algorithmic}
\usepackage{graphicx}
\usepackage{xcolor}
\usepackage{booktabs}
\usepackage{tabularx}

\usepackage{multirow}
\usepackage{subcaption}
\usepackage{array}

\definecolor{PerfUp}{RGB}{0,120,0}     
\definecolor{PerfDown}{RGB}{180,0,0}   
\definecolor{PerfBest}{RGB}{0,90,180} 

\newcommand{\up}{\textcolor{PerfUp}{$\uparrow$}}
\newcommand{\down}{\textcolor{PerfDown}{$\downarrow$}}

\newcommand{\best}{\textcolor{PerfBest}{$^\star$}}

\begin{document}

\title{Evaluating LLM Coding Agents on SZ-Family Lossy Compression Across Architectures}


\author{
\IEEEauthorblockN{Changqing Li$^{\dagger}$, Shouwei Gao$^{\dagger}$, Kai Zhao$^{*}$, Sheng Di$^{\ddagger}$, Wenqian Dong$^{\dagger}$}
\IEEEauthorblockA{
$^{\dagger}$Oregon State University, $^{*}$Florida State University, $^{\ddagger}$Argonne National Laboratory
}


}


\maketitle

\IEEEpubidadjcol
\input{sections/abstract}


\input{sections/introduction}

\input{sections/background}
\input{sections/evaluation}
\input{sections/conclusion}

\bibliographystyle{IEEEtran}
\bibliography{refs}

\end{document}

%% file: sections/abstract.tex

\begin{abstract}
Large language model (LLM) coding agents are increasingly applied to code translation and optimization, yet their effectiveness in performance-critical high-performance computing (HPC) settings remains poorly characterized. This paper evaluates LLM-based coding workflows on SZ-family error-bounded lossy compression kernels, which combine numerical constraints with memory-intensive and control-flow-heavy implementations. We study two representative CUDA workloads (SZp and SZx) and target two heterogeneous execution platforms: NVIDIA GPUs and Cerebras wafer-scale accelerators. Focusing on single-agent iterative generation, we analyze not only final throughput but also agent runtime behavior, including iteration patterns, sensitivity to prompt specification, and characteristic failure modes. Our results reveal a pronounced cross-architecture divergence. On GPUs, stronger models can achieve substantially higher throughput but exhibit increased sensitivity to prompt precision and optimization guidance, whereas on Cerebras the dominant challenge lies in producing runnable programs under a PE-centric spatial execution model. We further observe that LLM agents are more effective on modular kernels (SZx) than on tightly coupled bit-level pipelines (SZp), where structural dependencies hinder optimization progress. These findings suggest that evaluating LLM coding agents for HPC requires accounting for both performance outcomes and architecture-specific robustness, and that success on thread-based platforms does not directly transfer to spatial accelerators.
\end{abstract}

%% file: sections/introduction.tex
\section{Introduction}


Large language models (LLMs) have rapidly evolved from natural language
interfaces into practical programming assistants, demonstrating strong
capabilities in code synthesis, translation, and iterative refinement
across a wide range of software engineering tasks~\cite{chen2021codex,li2022competition}.
These advances have motivated growing interest in applying LLM-based
coding agents to system-level and performance-critical domains.

Beyond basic code generation, recent work has shown that LLM-based agents can translate programs across programming languages and platforms, and can even perform non-trivial performance optimization when guided by execution feedback~\cite{roziere2023code,llmcompiler2024}. These capabilities have sparked growing interest in applying AI coding agents to complex hardware and system software, where correctness, performance, and architecture-specific reasoning are all critical. 


A particularly challenging setting is heterogeneous high-performance
computing (HPC), where developers must simultaneously contend with
\emph{diverse programming languages} and \emph{architecture-specific
execution models}.
HPC software is typically written in low-level languages such as C/C++,
augmented with specialized programming models (e.g., CUDA for GPUs or
spatial DSLs for wafer-scale accelerators).
Achieving high performance requires careful reasoning about parallelism,
memory access patterns, and synchronization, and code optimized for one
architecture often cannot be directly ported to another without
substantial redesign.
These characteristics raise a natural question: \emph{how well do
current LLM-based coding agents perform when faced with fundamentally
different programming models and hardware architectures?}

In this work, we address this question through a focused evaluation
study.
Rather than proposing new agent designs, we empirically examine the
current capabilities and limitations of LLM-based coding agents on
realistic HPC workloads.
We select two SZ-family lossy compression codes as our evaluation target, SZp~\cite{zhao2022cuszp} and SZx~\cite{szx2023}.
Lossy compression is widely used in scientific computing to reduce data
movement and storage costs, and its performance directly impacts
end-to-end application throughput.
SZ-family compressors combine numerical computation, irregular control
flow, and memory-intensive bit-level kernels, making them representative
and challenging workloads for both code translation and performance
optimization.

\begin{figure}[t]
    \centering
    \includegraphics[width=0.8\columnwidth]{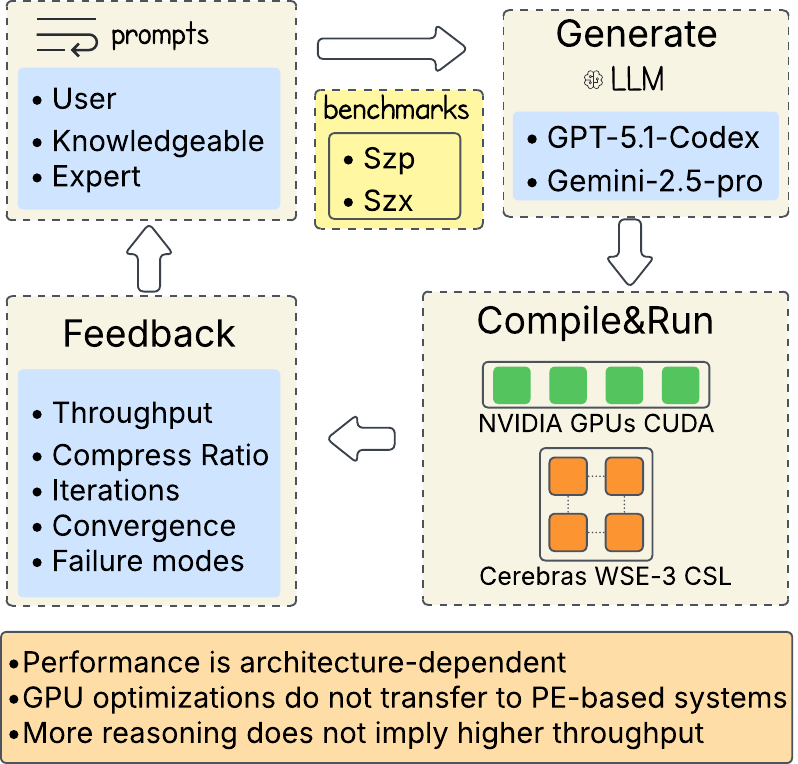}
      \caption{
    Overview of our evaluation framework.
    }
    \label{fig:overview2}
\end{figure}

We evaluate these workloads across two fundamentally different
architectural platforms: CUDA-enabled GPUs and Cerebras wafer-scale
accelerators.
GPUs employ a thread-based execution model with hierarchical memory,
whereas Cerebras systems rely on spatially distributed processing
elements and static dataflow organization.
This contrast allows us to assess whether coding agents that perform
well on familiar thread-based architectures can generalize to
architectures with radically different programming abstractions.
We compare different model families (i.e., GPT from OpenAI and Gemini from Google) and prompting tiers on final code performance. Figure~\ref{fig:overview2} provides an overview of our evaluation framework, illustrating the interaction between SZ-family benchmarks, agent workflows, target architectures, and analysis metrics that structure the remainder of this paper.
Our contributions are threefold. First, we present a performance-centric evaluation of LLM coding agents on realistic HPC workloads, emphasizing end-to-end throughput and compression efficiency rather than correctness alone. Second, we provide a cross-architecture analysis that contrasts agent behavior and performance on thread-based GPUs and PE-based wafer-scale accelerators, revealing qualitatively different success and failure modes. Third, we quantitatively characterize the runtime behavior of AI coding agents, including reasoning iterations, convergence patterns, and optimization trajectories, which offers new insights into how agentic LLM systems operate in performance-critical environments.



%% file: sections/background.tex
\section{Background}
\subsection{Error-Bounded Lossy Compression and the SZ Family}
Error-bounded lossy compression is widely used in HPC to reduce storage and I/O while guaranteeing user-specified absolute (or relative) error bounds \cite{lindstrom2014fixed}. Most scientific compressors follow a \emph{predict, quantize, and encode} pipeline: predictors exploit spatial or temporal locality, quantization enforces the error bound by binning residuals, and entropy (or byte)-level encoding removes redundancy \cite{tao2017significantly, di2016fast}. The SZ family is a representative design for floating-point fields \cite{di2014efficient}, evolving from CPU-centric Lorenzo-style predictors with entropy coding (SZ~1/2) \cite{di2016fast} to a modular, extensible architecture (SZ~3) that supports multiple predictors, quantizers, and encoders \cite{tao2022sz3}. GPU-focused variants (e.g., cuSZx/cuSZp) further restructure prediction, metadata construction, and encoding to match massive parallelism and memory-coalescing constraints \cite{li2020cusz, zhao2022cuszp}, making SZ-style codes a challenging and realistic target for cross-architecture translation due to their tight coupling between correctness (error guarantees) and low-level data layout/performance.

\subsection{LLM Agents for Code Generation}
Code-oriented LLMs have progressed from single-pass synthesis to \emph{agentic} workflows that iteratively plan, generate, compile or test, and refine code \cite{chen2021codex, yao2023react, wu2024autocoder}. More structured multi-agent designs decompose work into specialized roles (e.g., translator, optimizer, or validator) and can improve robustness on long-horizon tasks \cite{li2023camel, hong2024metagpt, openevolve}. However, most evaluations emphasize functional correctness (e.g., pass@k, unit tests) rather than architecture-aware performance, leaving open how well such agents handle HPC kernels that require careful memory hierarchy use, parallel decomposition, and portability-driven optimization.

\subsection{Cerebras Architecture}
Cerebras wafer-scale systems integrate a large 2D mesh of lightweight processing elements (PEs) with fast on-wafer communication \cite{feldman2021cerebras, lie2020wafer}. Unlike thread-based GPU models, computation is mapped \emph{spatially}: each PE runs a fixed program with explicit neighbor communication \cite{feldman2021cerebras, ho2022architecture}, and the placement or data movement is largely determined at compile time \cite{ho2022architecture, laurie2021programming}. This PE-centric, static mapping model shifts the burden to explicit reasoning about layout, synchronization, and communication \cite{laurie2021programming, cerebras-sdk}, providing a stringent test for LLM-driven translation beyond conventional CPU/GPU abstractions.

%% file: sections/evaluation.tex
\section{Experimental Setup}

\subsection{Benchmarks and Platforms}
We evaluate LLM-driven code translation and optimization using two SZ-family kernels, SZp~\cite{10.1145/3581784.3607048} and SZx~\cite{szx2023}, as representative error-bounded lossy compression workloads with CUDA validation version. Both kernels combine strict numerical error guarantees with throughput-critical GPU implementation details, including quantization logic and byte-/bit-level metadata handling, making them realistic targets for architecture-aware translation.

Experiments run on an NVIDIA V100 GPU and a Cerebras CS-3 system (WSE-3). On the GPU backend, we report end-to-end compression/decompression throughput (GB/s) for functionally correct implementations validated by our test suite. On the Cerebras backend, we report whether a runnable program is produced within a fixed iteration budget and, when successful, the execution time reported by the platform runtime; configurations that report zero execution time are treated as failures.

\subsection{Models and Agent Workflows}
We compare two LLMs, \texttt{GPT-5.1-Codex}~\cite{openai_codex} and \texttt{Gemini-2.5-pro}~\cite{gemini_report}, under two generation workflows.
The LLM agents iteratively generate and refine code in a compile/run feedback loop. The agent receives compiler diagnostics, runtime logs, and validation outcomes, and revises the implementation until it passes validation or reaches a fixed iteration budget. Each single-agent instance is equipped with a Model Context Protocol (MCP) environment that supports file I/O and command execution (e.g., compilation and benchmarking), enabling end-to-end iteration with execution feedback.


\subsection{Prompt Tiers}
To quantify sensitivity to user guidance, we define three prompt tiers. (i) The \textit{user} tier specifies the goal and constraints with the most fundamental domain detail. (ii) The \textit{knowledgeable} tier adds concise domain context (e.g., SZ pipeline stages and error-bound requirements) while keeping instructions lightweight. (iii) The \textit{expert} tier further adds architecture-aware guidance, such as memory-coalescing and parallel decomposition hints on GPU, or spatial mapping and explicit communication considerations on Cerebras. Unless otherwise noted, all tiers use the same inputs, validation rules, and iteration budgets.

\subsection{Evaluation Metrics}
For the GPU backend, we report (i) throughput (GB/s) for compression and decompression speed on those implementations that can pass correctness validation. We also track agent behavior, including (ii) the number of iterations required to reach a correct implementation and (iii) per-configuration success rates. For the Cerebras backend, the primary outcome is (iv) runnability (compile and execute within budget), along with (v) the execution time reported by the runtime for successful runs.

\section{Single-Agent Performance}

\subsection{CUDA Code Translation Performance on Nvidia V100}

\subsubsection{Throughput comparison across agents}
\begin{figure}[!htb]
  \centering
  \includegraphics[width=\columnwidth]{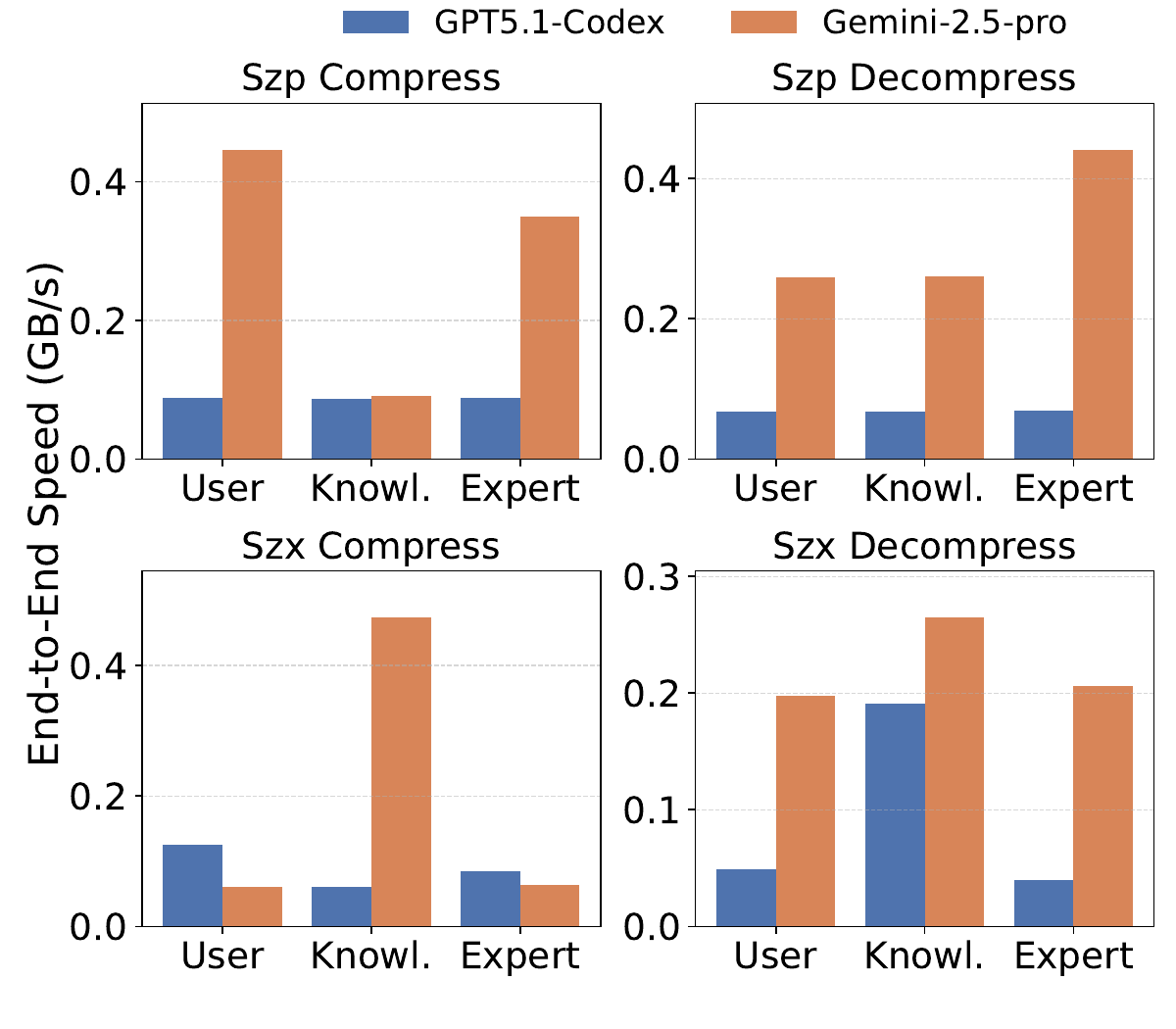}
  \caption{Throughput performance of SZp and SZx on Nvidia V100 (Knowl.=Knowledgeable).}
  \label{fig:singleagent-throughput}
\end{figure}

Figure~\ref{fig:singleagent-throughput} reports end-to-end throughput on V100 for SZp and SZx across two LLMs and three prompt tiers. Across both kernels, \texttt{Gemini} achieves higher peak throughput than \texttt{GPT}, reaching 0.446~GB/s (compression) and 0.441~GB/s (decompression) on SZp, versus 0.084-0.088~GB/s and 0.04-0.069~GB/s for \texttt{GPT}. Similar gaps appear on SZx, where \texttt{Gemini} peaks at 0.474~GB/s compression compared to \texttt{GPT} at 0.126~GB/s. The two models differ sharply in prompt sensitivity: \texttt{Gemini} exhibits large tier-to-tier variance (up to about 5$\times$ on SZp compression), while \texttt{GPT} remains largely prompt-invariant (within 5\% on SZp compression). Finally, iteration count does not reliably predict performance: \texttt{GPT} often uses more iterations (e.g., up to 41) but produces slower code, indicating that additional refinement is frequently spent on achieving correctness or resolving compilation or runtime issues rather than discovering bandwidth-critical optimizations.

\begin{figure}[!htb]
  \centering
  \includegraphics[width=\linewidth]{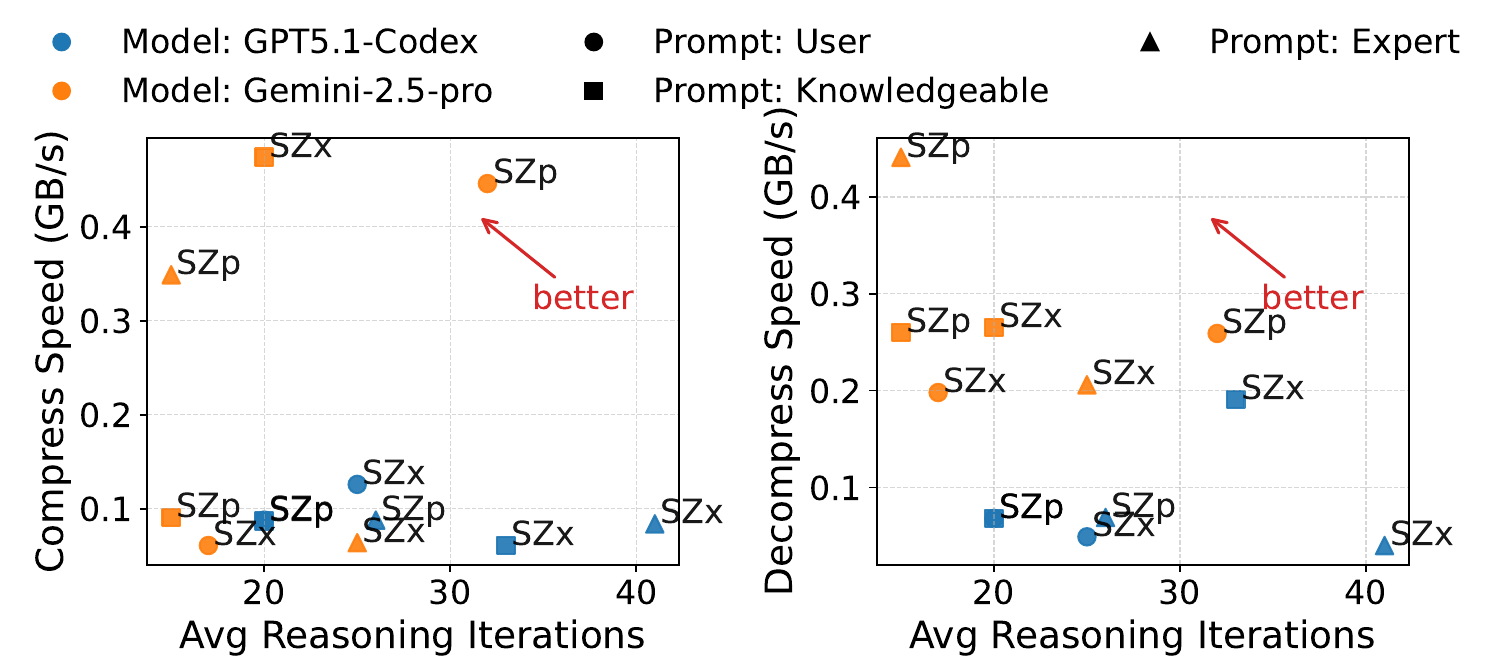}
  \caption{Throughput vs. reasoning iterations.}
  \label{fig:reasoning-vs-throughput}
\end{figure}

\subsubsection{Throughput v.s. reasoning iterations}
Figure~\ref{fig:reasoning-vs-throughput} further validates that deeper reasoning traces do not reliably translate into higher throughput. Although \texttt{GPT} typically uses more iterations (20-41 for SZp and 25-41 for SZx) than \texttt{Gemini} (15-32 for SZp and 17-25 for SZx), it rarely reaches the high throughput region. For example, \texttt{GPT} at 41 iterations (SZx Expert) delivers only 0.084~GB/s compression and 0.04~GB/s decompression, whereas \texttt{Gemini} achieves substantially higher throughput with fewer iterations, reaching 0.349~GB/s compression and 0.441~GB/s decompression on SZp at 15 iterations. Overall, the strongest performance differences align more with model choice and prompt tier (\texttt{Gemini} exhibits large tier-to-tier variation) than with iteration evolution, suggesting that additional iterations are often spent on correctness and debugging rather than discovering bandwidth-critical optimizations.

\textbf{Takeaways:} (i) \texttt{Gemini} is prompt-sensitive, whereas \texttt{GPT} is comparatively prompt-robust; (ii) Increasing reasoning iterations does not reliably improve system-level optimizations (or end-to-end throughput).

\subsection{Cerebras Specific Language on WSE-3}

Table~\ref{tab:cerebras-perform} summarizes single-agent outcomes on CS-3/WSE-3. \texttt{GPT} produces runnable CS-3 code consistently across prompts, while \texttt{Gemini} fails across all prompts; reasoning-iteration counts vary widely but do not predict successful execution. The primary challenge is not performance tuning but reaching a runnable program: most failures come from incorrect host-to-PE data movement and violations of Cerebras’ static spatial mapping and communication requirements, which typically surface as kernel stalls or `zero' reported in execution time. Under the same iteration budget, \texttt{GPT} is more reliable, producing runnable implementations across prompt tiers and exhibiting largely prompt-invariant behavior once a valid CSL structure is established. In contrast, \texttt{Gemini} frequently fails to produce runnable code within the budget, and higher-tier prompts do not consistently improve its success rate.

\textbf{Takeaways:} (i) On CS-3/WSE-3, correctness dominates; (ii) Success hinges on satisfying Cerebras’ static dataflow/mapping and host–PE movement constraints, not on deeper reasoning or prompt “expertise.” (iii) GPT reliably reaches runnable CSL across prompts, while Gemini often fails within the same budget.

\begin{table}[t]
\centering
\small
\setlength{\tabcolsep}{3pt}        
\renewcommand{\arraystretch}{1.05} 
\caption{LLM-based code generation on Cerebras CS-3}
\label{tab:cerebras-perform}
\begin{tabularx}{\columnwidth}{>{\raggedright\arraybackslash}X l l c c}
\toprule
Model & Task & Prompt & Reasoning & Exec. Time (s) \\
\midrule
\multirow{6}{*}{GPT-5.1-Codex}
 & Szp & User & 74 & 134 \\
 & Szp & Knowledgeable & 100$^{\dagger}$ & 134 \\
 & Szp & Expert & 79 & 134 \\
 \cmidrule(lr){2-5}
 & Szx & User & 100$^{\dagger}$ & 134 \\
 & Szx & Knowledgeable & 99 & 134 \\
 & Szx & Expert & 100$^{\dagger}$ & 134 \\
\midrule
\multirow{6}{*}{Gemini-2.5-pro}
 & Szp & User & 100$^{\dagger}$ & 0$^{\ddagger}$ \\
 & Szp & Knowledgeable & 37 & 0$^{\ddagger}$ \\
 & Szp & Expert & 100$^{\dagger}$ & 0$^{\ddagger}$ \\
 \cmidrule(lr){2-5}
 & Szx & User & 9 & 0$^{\ddagger}$ \\
 & Szx & Knowledgeable & 11 & 0$^{\ddagger}$ \\
 & Szx & Expert & 12 & 0$^{\ddagger}$ \\
\bottomrule
\end{tabularx}
\vspace{0.35em}
\footnotesize
\noindent
$^{\dagger}$ Reasoning budget exhausted. \quad
$^{\ddagger}$ Fail to generate runnable execution.
\end{table}

\section{Deep-Dive of A Step-by-Step Analysis}

\begin{figure}[t]
  \centering
  \includegraphics[width=\linewidth]{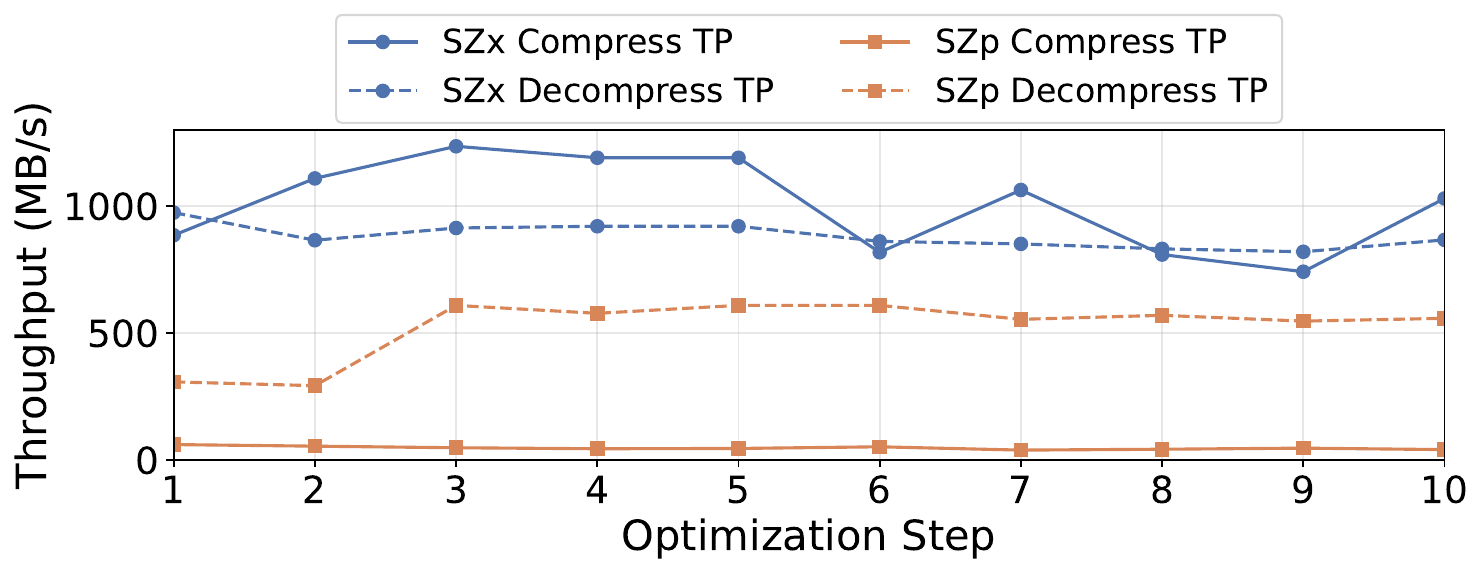}
  \caption{Throughput evolution across optimization steps for Cu-SZx and Cu-SZp on a 25\,MB dataset.}
  \label{fig:singleagent_trajectory}
\end{figure}

To move beyond a single statistic value of throughput, we analyze how the single-agent workflow improves CUDA code through iterative refinement, using intermediate code revisions, tool logs, and execution feedback. Across runs, the agent repeatedly applies a small set of high-leverage transformations, including kernel fusion, migrating CPU-side stages to the GPU, restructuring memory access patterns, and simplifying control flow while preserving error-bound correctness.

\subsection{Evolution of Throughput Performance} Figure~\ref{fig:singleagent_trajectory} shows trajectory-level behavior on a fixed 25\,MB input, tracking throughput and compression ratio across successive optimization steps for Cu-SZx and Cu-SZp. Two observations emerge. First, tuning is non-monotonic: iterations that improve one metric can regress another, especially when changes touch packing/metadata kernels. Second, SZx and SZp induce qualitatively different optimization dynamics, reflecting their adaptive-ratio versus fixed-ratio characteristics.

\subsection{Optimization Trajectories}
Table~\ref{tab:singleagent_trace_tp} and Table~\ref{tab:szp_trace_tp} makes these dynamics concrete by reporting median compression or decompression throughput ($T_c$, $T_d$) and compression ratio at each step. Early iterations (i.e., {Step 2-5 in SZp and Step 2-6 in SZX}) tend to target system-level bottlenecks that dominate end-to-end cost, such as eliminating host-to-device round trips, moving prefix-sum/offset bookkeeping and pack/unpack stages onto the GPU, and fusing kernels to reduce launch overhead and intermediate global-memory traffic. These changes account for most of the sustained throughput gains, indicating that the agent can reliably identify the highest-impact CUDA bottlenecks.

Later iterations (i.e.,{Step 7-12 in SZp and Step 7-10 in SZX}) shift toward micro-kernel edits (e.g., warp-level primitives, aligned/vectorized stores, and more aggressive parallel bit packing). With this, performance becomes highly sensitive to synchronization, thread cooperation, and coalesced writes. As a result, the marginal gains diminish, and regressions become more common: seemingly reasonable edits can increase synchronization pressure or create uncoalesced stores that offset the intended benefit.

As shown in Table~\ref{tab:szp_trace_tp}, the gap to mature hand-optimized implementations (e.g., cuSZp) is most pronounced in the bit-level packing kernels. Expert code tightly controls synchronization and keeps packing operations within warp- or block-local scopes, preserving memory coalescing and bandwidth. By contrast, LLM-generated variants more often introduce serial regions, unnecessary synchronization  (i.e.,{Step 7}), or inefficient store patterns (i.e., {Step 10}), which lowers effective bandwidth and throughput.

Overall, the LLM agent exhibits strong strategic optimization behavior, which consistently identifies and removes major bottlenecks, but weaker tactical CUDA craftsmanship in sensitive bit-level kernels. Practically, this suggests a two-level workflow: use agents to propose high-level restructuring, but validate late-stage micro-optimizations with profiling signals such as achieved bandwidth, global-store efficiency, occupancy, and synchronization overhead to avoid over-tuning that leads to regressions.

\textbf{Takeaways.} Most stable gains come from early, system-level fixes (GPU-ifying stages, removing H2D/D2H round trips, kernel fusion). Late-stage micro-kernel tweaks, especially bit-packing/metadata, are brittle and should be profile-gated (bandwidth, store efficiency, occupancy, sync stalls) to avoid regressions.

\begin{table}[!t]
\centering
\caption{Cu-SZx optimization trace. 
Arrows indicate throughput change from the previous step:
\textcolor{PerfUp}{$\uparrow$} increase,
\textcolor{PerfDown}{$\downarrow$} decrease.
$\star$ marks the best value within each trace.}
\label{tab:singleagent_trace_tp}
\scriptsize
\setlength{\tabcolsep}{4pt}
\begin{tabularx}{\columnwidth}{c r r r p{3.2cm}}
\toprule
Step & $T_c$ & $T_d$ & Ratio & Action \\
\midrule
1  & 886.0             & 975.2\best        & 3.50 & Baseline \\
2  & 1110.0\,\up       & 866.3\,\down      & 4.75 & Initial GPU parallelism \\
3  & 1236.7\,\up       & 914.5\,\up        & 4.75 & Zstd backend \\
4  & 1191.4\,\down     & 921.5\,\up        & 4.75 & Kernel parallelization \\
5  & 1191.4            & 921.5             & 4.75 & GPU prefix-sum (offsets) \\
6  & 818.7\,\down      & 861.8\,\down      & 2.15 & Remove XOR prefix; LZ-byte encoding; GPU scan \\
7  & 1064.4\,\up       & 851.8\,\down      & 4.75 & Remove host radius copy \\
8  & 809.4\,\down      & 832.3\,\down      & 2.15 & Chunked pipeline \\
9  & 742.2\,\down      & 820.7\,\down      & 2.15 & Dual-stream chunk pipeline \\
10 & 1031.0\,\up       & 867.7\,\up        & 4.75 & Reduce CPU-side work \\
11 & 979.0\,\down      & 846.8\,\down      & 4.75 & GPU O-array generation \\
12 & 1270.6\,\up\best  & 911.2\,\up        & 7.99 & Block size tuning \\
\bottomrule
\end{tabularx}
\end{table}

\begin{table}[!t]
\centering
\caption{Cu-SZp optimization trace.}
\label{tab:szp_trace_tp}
\scriptsize
\setlength{\tabcolsep}{4pt}
\begin{tabularx}{\columnwidth}{c r r r p{3.2cm}}
\toprule
Step & $T_c$ & $T_d$ & Ratio & Action \\
\midrule
1  & 60.8\best        & 307.9              & 15.87 & Baseline \\
2  & 54.5\,\down      & 292.9\,\down       & 15.87 & CUB scan; CUDA streams \\
3  & 48.2\,\down      & 609.4\,\up         & 15.87 & GPU bit pack/unpack \\
4  & 44.3\,\down      & 578.2\,\down       & 15.87 & Grid-stride packing \\
5  & 45.4\,\up        & 609.5\,\up         & 15.87 & Kernel fusion; remove intermediates \\
6  & 52.0\,\up        & 609.6\,\up\best    & 15.87 & Parallel byte-count writes \\
7  & 38.9\,\down      & 554.4\,\down       & 15.87 & Parallel remainder-bit pack \\
8  & 42.4\,\up        & 570.3\,\up         & 15.87 & Pre-alloc output buffer \\
9  & 46.7\,\up        & 547.4\,\down       & 15.87 & Warp-level ballot pack \\
10 & 40.9\,\down      & 558.5\,\up         & 15.87 & Vectorized aligned writes \\
\bottomrule
\end{tabularx}
\end{table}

%% file: sections/conclusion.tex
\section{Conclusions}

This paper evaluates LLM coding agents on SZ-family lossy compression across heterogeneous architectures, measuring end-to-end throughput and optimization trajectories. We find: (i) GPU performance depends strongly on the LLM model and its interaction with prompting/feedback: some settings reach high throughput but are prompt-sensitive, while others are steadier but slower; (ii) cross-architecture robustness diverges, as WSE-3 correctness hinges on spatial mapping and host–PE data movement where GPU-strong models can fail; and (iii) kernel structure governs optimization, with agents improving modular SZx more reliably than tightly coupled SZp with bit-level packing. Overall, HPC agent evaluation should go beyond correctness to include throughput, convergence, and architecture-specific failure modes.  Our future work will expand kernels or hardware and explore hybrid agent and analysis workflows.

\section{Acknowledgment}

This work was supported by the U.S. Department of Energy, Office of Science, Advanced Scientific Computing Research (ASCR), under contract DE-AC02-06CH11357. This work is supported by the U.S. National Science Foundation (2514351, 2505118, 2514036, and 2513768). We thank the anonymous reviewers for their valuable feedback.